\documentclass[aps, pra, preprint]{revtex4-1}
\usepackage[utf8]{inputenc}
\usepackage[T1]{fontenc}
\usepackage[english]{babel}
\usepackage[dvipsnames]{xcolor}
\usepackage{amsmath, mathtools}
\usepackage{amssymb}
\usepackage{setspace}
\usepackage{graphicx}
\begin{document}
\title{Structural relaxation in quantum supercooled liquids: A mode-coupling approach}
\author{ Ankita Das$^{*}$, Eran Rabani$^{\dagger}$, Kunimasa Miyazaki$^{\ddagger}$, and Upendra Harbola$^{*}$}
\affiliation{ $^{*}$Inorganic and Physical Chemistry, Indian Institute of Science, Bangalore 560012, India.}
\affiliation{$^{\dagger}$ Department of Chemistry, University of California, and Materials Sciences Division, Lawrence Berkeley 
National Laboratory, Berkeley, California 94720, United States;\\ The Sackler Center for Computational Molecular and Materials Science, 
Tel Aviv University, Tel Aviv 69978, Israel.  }
\affiliation{$^{\ddagger}$Department of Physics, Nagoya University, Nagoya 464-8602, Japan}


\begin{abstract}
  We study supercooled dynamics in quantum hard-sphere liquid using quantum mode-coupling formulation. 
  In the moderate quantum regime, classical cage effects lead to slower dynamics compared to strongly quantum 
  regime, where tunneling overcomes classical caging, leading to faster relaxation. As a result, the glass transition 
  critical density can become significantly higher than for the classical liquids. Perturbative approach is used to solve time 
  dependent quantum mode-coupling equations to study in detail the dynamics of the supercooled liquid in moderate 
  quantum regime. Similar to the classical case, relaxation time shows power-law increase with increasing density
  in the supercooled regime. However, the power-law exponent is found to be dependent on the quantumness;
  it increases linearly as the quantumness is increased in the moderate quantum regime.
\end{abstract}

\maketitle

\section{Introduction}
Almost every liquid undergoes a glass transition when supercooled below its freezing temperature, bypassing the formation of crystalline state. 
The time scale for structural relaxation in the supercooled liquid increases very rapidly on cooling \cite{2013i}. Thus supercooled liquids are 
characterized by high shear viscosity. Experimentally, the glass transition is assumed to take place when viscosity is 
$O(10^{13}$ poise) \cite{2013i,angel_plot_ref}.

Mode-coupling theory (MCT) is a generalized hydrodynamic approach to study supercooled dynamics in simple liquids \cite{Kim-Mazenko}.
MCT deals with structural dynamics of liquids encoded in the time evolution of density correlation functions \cite{mct_1992, das_2011, MCT2005, gotze_mct}.
The important result of  MCT consists of a self-consistent expression for generalized transport coefficients in terms of slowly decaying density
correlations. The  theory has been applied successfully to study liquid-glass transition in one- and two-component 
liquids \cite{das-mazenkoPRA1986, GotzeJPhysC1984, uh-das-PRE2002, BossePRA1991}.   
 The relation between structural relaxation in supercooled liquids and frequency-dependent specific heat has been 
 studied \cite{Harbola2001} using mode-coupling theory(MCT). 

Most of the work on glass transition has focused on the classical regime of liquid state behavior, where the de Broglie thermal wavelength associated 
with a particle is significantly smaller than the particle size. As most of the glass forming liquids fall in the classical regime, it is clear that classical 
approximation is generally justified. However, there are several important and interesting examples, such as superfluid helium under high pressure, where 
quantum fluctuation and glassiness coexist. There are clear deviations from classical predictions, for example, temperature dependence of specific heat at 
low temperature ($~ 1-10$ K) which has been attributed to quantum tunneling \cite{zeller_paul_1971, ref_low_sph, 2019_sph}. In any case, at low 
temperatures one expects for quantum features to show up. 

To explore quantum effects on supercooled dynamics, MCT has been extended to quantum 
regime known as, quantum mode-coupling theory (QMCT) \cite{R2002}. As intuition may suggest that large quantum fluctuations serve to inhibit glass 
formation as tunneling and zero-point energy allow   particles   to   traverse barriers and facilitate movement. But interestingly, both ring-polymer 
molecular dynamics (RPMD) and QMCT indicate that the dynamical phase diagram of glassy quantum fluids is reentrant; low quantum fluctuations 
promote the glass transition and further increase in quantum fluctuations leads to inhibition to glass formation \cite{Markland2011}. 
 
In the present work our goal is to study quantum effects on the dynamics in supercooled state of a quantum liquid. We present a detail analysis of the 
dynamics in the moderate quantum regime where the quantum effects assist in forming the glass phase, as opposed to the strong quantum regime where 
quantum fluctuations tend to avoid the glass phase \cite{Markland2011}. In the next section, we describe essential features of quantum dynamics of density-correlation near liquid-glass transition using QMCT.

\section{Basic equation of QMCT}

QMCT is a hydrodynamic theory \cite{Markland2012,R2002} which is based on the dynamics of hydrodynamic variables $\{\rho,$g$\}$, where 
$\rho$ is the mass density, $g$ is the momentum density. The dynamics of the density fluctuations in supercooled liquids and the subsequent structural 
arrest provides an idea of the glass transition. 

Prime goal of the theory is to describe time-evolution of quantum density correlation defined as
\begin{equation}
C_{\rho\rho}(q,t)=\frac{1}{N}\langle\hat\rho(-q,0)\hat\rho(q,t)\rangle
\end{equation}
where $\langle ... \rangle$ denotes quantum mechanical ensemble average and the density operator is $\hat\rho(q)=\sum_{i=1}^{N}e^{iq\cdot \hat r_i}$, $N$ 
being the total number of particles, $r_i$ is position of the $i^{th}$ particle and $q$ is the wave-vector. The zero-time quantum density correlation is the 
quantum structure factor, denoted by $S(q)$. It is a difficult task to evaluate $C_{\rho \rho}(q,t)$ directly with full quantum mechanical evolution. 
Ring polymer approach, which maps quantum particles to closed classical polymers \cite{Chandler1981}, allows us to use semi-classical tools to evaluate 
quantum correlations. For this, one defines Kubo transformed quantities. The Kubo transformed density correlation is defined as
\begin{equation}
 \tilde C_{\rho\rho}(q,t)=\frac{1}{N\beta \hbar}\int_0^{\beta \hbar}d\lambda\langle\hat\rho(-q,0)\hat\rho(q,t+i\lambda)\rangle,
\end{equation}
where $\beta=\frac{1}{k_BT}$ is the inverse temperature, $k_B$ is Boltzmann  constant and $\hbar$ is the Planck's constant. The notation $\tilde A$ 
implies Kubo-transformed quantity, defined as  $\tilde A(q)=\frac{1}{\beta \hbar}\int_0^{\beta \hbar}d\lambda e^{-\lambda \hat H}A(q)e^{\lambda \hat H}$, 
where $\hat H$ is the hamiltonian of the system. The zero-time Kubo-transformed density correlation is denoted by $\tilde S(q)$, known as Kubo-transformed 
structure factor.

The projection operator formalism \cite{2013i,Z2001,MCT2005} of Zwanzig and Mori is used to develop the QMCT to study time-dependence of 
$\tilde{C}_{\rho\rho}$. The projection operator is defined as $\hat P=\sum_q\frac{|\tilde A(q)>< A(-q)|}{< A(-q)\tilde A(q)>}$ , where $A(q)$ is 
the  vector $(\hat\rho(q),\hat g_z(q))^{T}$, and $g_z$ is the longitudinal component of momentum density. 

Following the procedure outlined in Ref. \cite{R2002, Markland2012}, the following QMCT equation for $\tilde{C}_{\rho\rho}$ is obtained
\begin{equation}
 \frac{d^2{\tilde{ C}}_{\rho \rho}(q,t)}{dt^2}+\Omega_q^2 \tilde C_{\rho \rho}(q,t) +\int_0^t dt^{\prime} M(q,t-t^{\prime})
 \frac{d {\tilde{ C}}_{\rho \rho}(q,t^{\prime})}{dt^{\prime}}=0
 \label{1}
  \end{equation}
  where $\Omega_q^2=\frac{q^2}{m\beta\tilde S(q)}$ is the frequency term, $m$ is mass of the particle and 
  $M(q,t)=\frac{1}{2\pi}\int_{-\infty}^{\infty}dw e^{-i\omega t}M(q,\omega)$ with
  
 \begin{equation}
 M(q,\omega)=\frac{\hbar m\beta^2}{16\omega\pi^3q^3n}\int_0^{\infty} dk k\int_{|q-k|}^{q+k} d\kappa \kappa V^2_{q,k,\kappa}
 \int_{-\infty}^{\infty}d\omega^\prime \Phi_{\kappa}(\omega^\prime)\Phi_k(\omega-\omega^\prime)\omega^\prime(\omega-\omega^\prime)T(\omega^\prime,\omega-\omega^\prime)
 \label{2}
 \end{equation}
  where $\Phi_q(\omega)=\int_{-\infty}^{\infty}dte^{i\omega t}\tilde C_{\rho \rho}(q,t)$ is the Fourier transform of the density-density, Kubo transformed, correlation function.
 
  The last term in Eq.(\ref{1})  is (quantum) mode-coupling nonlinear term where  $M(q,t)$ is the memory function which originates due to correlation of 
  random-forces. The  memory function is  approximated by decoupling four-point density correlation in terms of product of two-point correlation 
  functions and the projected dynamics of the random-force is replaced with the full dynamics projected onto the slow decaying modes. 
  The  approximated vertex function is given by \cite{Markland2012}
 \begin{equation}
 V_{q,k,\kappa}=\frac{\Delta n(\Omega_{\kappa})\Delta n(\Omega_k)\zeta_{q,k,\kappa}}{S({\kappa})S(k)K(\Omega_{\kappa},\Omega_k)}
 \left[\frac{(\Omega_k+\Omega_{\kappa})^2-\Omega_q^2}{\Omega_k+\Omega_{\kappa}} \
\right]
 \end{equation}
  where
\begin{equation}
 \zeta_{q,k,\kappa}=\frac{\Omega_qS(q)S(k)S({\kappa})-\frac{\hbar \Delta n(
 \Omega_q)}{4m}[(q^2+k^2-\kappa^2)S(\kappa)+(q^2-k^2+\kappa^2)S(k)]}{\Omega_q\Delta n(\Omega_k+\Omega_{\kappa})-
 (\Omega_k+\Omega_{\kappa})\Delta n(\Omega_q)},
 \label{eq5}
\end{equation}

 \begin{equation}
  K(\Omega_{\kappa},\Omega_k)=\frac{T(\Omega_{\kappa},\Omega_k)}{\Omega_{\kappa}+\Omega_k}
  +\frac{T(-\Omega_{\kappa},\Omega_k)}{\Omega_{\kappa}-\Omega_k}.
  \label{eq6}
 \end{equation}
Here $n$ is the number density, $n(w)=\frac{1}{e^{\beta\hbar w}-1}$ is the Bose distribution, $\Delta n(w)=n(w)-n(-w)$ and 
$T(w_1,w_2)=n(-w_1)n(-w_2)-n(w_1)n(w_2)$.

In order to solve Eq.(\ref{1}), we first need to evaluate the memory function in the frequency-domain which requires the pre-knowledge 
of the whole set of density correlations in the frequency-domain as given in Eq.(\ref{2}). The numerical calculation of $M(q,\omega)$ 
involves calculation of several integrals one inside another, which is computationally costly. Once the frequency-dependent memory function is 
obtained using some proper guess value for $\Phi_q(\omega)$, the memory function can be Fourier transformed to time-domain to obtain 
$\tilde{ C}_{\rho \rho}(q,t)$. The new set of $\Phi_q(\omega)$ can be calculated through Fourier-transforming $\tilde{ C}_{\rho \rho}(q,t)$ 
and the memory function can be updated in self-consistent manner. These  numerical calculations require back and forth frequency and time-domain 
transformations, where a small error  may add up in successive self-consistent loops and lead to a wrong result or may not converge at all. These 
difficulties can be avoided using perturbative approach to solve the dynamics of quantum density-correlation as we describe below.

\subsection{Perturbative approach}
 Equation(\ref{2}) contains a convolution of products of $\phi(q,\omega)$ with $T(\omega,\omega^{\prime})$ which makes it difficult to 
 analytically transform the memory-term $M(q,\omega)$ in the time-domain . 
 The function $\frac{1}{\omega}{\omega^\prime(\omega-\omega^\prime)T(\omega^\prime,\omega-\omega^\prime)}$ present in the 
 memory function can be expanded around $\beta \hbar$ as
 \begin{equation}
\begin{multlined}
 \frac{1}{\omega}\omega^\prime(\omega-\omega^\prime)T(\omega^\prime,\omega-\omega^\prime)= 
 \\ \frac{1}{\beta\hbar}+\frac{\beta\hbar}{12}\omega^{\prime}(\omega-\omega^{\prime})-
 \frac{(\beta\hbar)^3}{720}\sum_{i=0}^{n=2}(-1)^i{\omega^{\prime}}^{i+1}{(\omega-\omega^{\prime}})^{n-i+1}
 \\ +\frac{(\beta\hbar)^5}{30240}\sum_{i=0}^{n=4}(-1)^i{\omega^{\prime}}^{i+1}{(\omega-\omega^{\prime}})^{n-i+1}
  -\frac{(\beta\hbar)^7}{1209600}\sum_{i=0}^{n=6}(-1)^i{\omega^{\prime}}^{i+1}{(\omega-\omega^{\prime}})^{n-i+1} 
  \\+\frac{(\beta\hbar)^9}{47900160}\sum_{i=0}^{n=8}(-1)^i{\omega^{\prime}}^{i+1}{(\omega-\omega^{\prime}})^{n-i+1}
  -\frac{691(\beta\hbar)^{11}}{210\times 13!}\sum_{i=0}^{n=10}(-1)^i{\omega^{\prime}}^{i+1}{(\omega-\omega^{\prime}})^{n-i+1}
  \\+O(13).
\end{multlined}
\end{equation}
Plugging this into the memory function expression, Eq.(\ref{2}), we get
\begin{equation}
\begin{multlined}
 M(q,\omega)=\frac{\hbar m\beta^2}{16\pi^3q^3n}\int_0^{\infty} dk k\int_{|q-k|}^{q+k} d\kappa \kappa 
 V^2_{q,k,\kappa}\int_{-\infty}^{\infty}d\omega^\prime \phi_{\kappa}(\omega^\prime)\phi_k(\omega-\omega^\prime)
 \\ \times \left[\frac{1}{\beta\hbar}+\frac{\beta\hbar}{12}\omega^{\prime}(\omega-\omega^{\prime})-
 \frac{(\beta\hbar)^3}{720}\sum_{i=0}^{n=2}(-1)^i{\omega^{\prime}}^{i+1}{(\omega-\omega^{\prime})}^{n-i+1}+...\right].
 \end{multlined}
 \label{3}
 \end{equation}
 
 Fourier transforming Eq.(\ref{3}) in the time-domain we obtain
\begin{equation}
 \begin{multlined}
 M(q,t)=\frac{ m\beta}{8\pi^2q^3n}\int_0^{\infty} dk k\int_{|q-k|}^{q+k} d\kappa \kappa V^2_{q,k,\kappa}
\times \Big[\tilde C_{\rho\rho}(\kappa,t)\tilde C_{\rho\rho}(k,t)
 \\ -\frac{(\beta\hbar)^2}{12}{{\tilde C}}_{\rho\rho}^{(1)}(\kappa,t){\tilde{ C}}_{\rho\rho}^{(1)}(k,t) -
 \frac{(\beta\hbar)^4}{720}\sum_{i=0}^{n=2}(-1)^i\tilde C_{\rho\rho}^{(i+1)}(\kappa,t)\tilde C_{\rho\rho}^{(n-i+1)}(k,t)-...\Big],
 \end{multlined}
 \label{4}
 \end{equation}
 where $\tilde C_{\rho\rho}^{(i)}(q,t)$ is the $i^{th}$ time derivative of density-correlation function.

In the following calculations the memory function in Eq. (\ref{4}) is computed perturbatively to the $12^{th}$ power of 
$\beta\hbar$ and the higher power terms  are ignored. We have
\begin{equation}
 \begin{multlined}
 M(q,t)\approx \frac{ m\beta}{8\pi^2q^3n}\int_0^{\infty} dk k\int_{|q-k|}^{q+k} d\kappa \kappa V^2_{q,k,\kappa}
 \times \Big[\tilde C_{\rho\rho}(\kappa,t)\tilde C_{\rho\rho}(k,t)
\\-\frac{{\Lambda^*}^2\tau^2}{12}{{\tilde C}}_{\rho\rho}^{(1)}(\kappa,t){\tilde{ C}}_{\rho\rho}^{(1)}(k,t) 
 -\frac{{\Lambda^*}^4\tau^4}{720}\sum_{i=0}^{n=2}(-1)^i\tilde C_{\rho\rho}^{(i+1)}(\kappa,t)\tilde C_{\rho\rho}^{(n-i+1)}(k,t)
 \\ -\frac{{\Lambda^*}^6\tau^6}{30240}\sum_{i=0}^{n=4}(-1)^i\tilde C_{\rho\rho}^{(i+1)}(\kappa,t)\tilde C_{\rho\rho}^{(n-i+1)}(k,t)
 -\frac{{\Lambda^*}^8\tau^8}{1209600}\sum_{i=0}^{n=6}(-1)^i\tilde C_{\rho\rho}^{(i+1)}(\kappa,t)\tilde C_{\rho\rho}^{(n-i+1)}(k,t)
 \\-\frac{{\Lambda^*}^{10}\tau^{10}}{47900160}\sum_{i=0}^{n=8}(-1)^i\tilde C_{\rho\rho}^{(i+1)}(\kappa,t)
 \tilde C_{\rho\rho}^{(n-i+1)}(k,t)-\frac{691{\Lambda^*}^{12}\tau^{12}}{210\times 13!}\sum_{i=0}^{n=10}(-1)^i\tilde C_{\rho\rho}^{(i+1)}(\kappa,t)
 \\ \times\tilde C_{\rho\rho}^{(n-i+1)}(k,t)\Big],
 \end{multlined}
 \label{5}
 \end{equation}
 where $\Lambda^*=\frac{\hbar}{\sqrt{m\sigma^2K_BT}}$ is the ratio of thermal wavelength with the particle size, $\tau=\sqrt{m\sigma^2\beta}$, 
 $\sigma$ is the diameter and $m$ is the mass of the particle. At $\Lambda^* \to 0$ limit the quantum vertex $V_{q,k,\kappa}$ exactly reduces 
 to the classical one \cite{gotze_84} , thereby in this limit quantum and classical memory function become equal. With gradual increase in $\Lambda^*$ the quantum 
 fluctuations becomes important and the quantum effect is well captured by the perturbative terms up to a certain value of $\Lambda^*$.

In the perturbative approach we can evaluate the memory function in the time-domain and Eq.(\ref{1}) can be solved in self-consistent manner 
bypassing back and forth Fourier-transforms to frequency and time-domain. Having lesser number of integrals in the time-dependent memory function 
makes it computationally less costly compared to calculating the memory term in frequency-domain. In this method the Kubo-transformed and quantum 
static structure factor are used as input. The quantum static structure factor for hard-sphere system is calculated using reference interaction site model (RISM), 
where the Percus-Yevick (PY) approximation is used as the closure equation. The Kubo-transformed structure factor is calculated using the approximate 
relation ($\tilde S(q)\approx \frac{2S(q)}{\beta\hbar\Delta n(\Omega_q)} $) \cite{Markland2012}. 

\subsection{Nonergodic parameter}
For a liquid the density-correlation decays in time and vanishes in the long time limit. As the liquid is supercooled, the density fluctuations become correlated 
over much longer time-scales. MCT predicts a sharp transition to the glassy state where these correlations are frozen at all times at all length-scales. 
The system is said to be trapped in a metastable state and therefore become nonergodic. The density fluctuations at $t\to \infty$, also referred to as 
nonergodic parameter, identifies this sharp transition. The nonergodic parameter is defined as the normalized density correlation at $t \to\infty$, 
\begin{equation}
 f_q=\frac{\tilde C_{\rho\rho}(q,t\to \infty)}{\tilde C_{\rho\rho}(q,t=0)}
 =\frac{\tilde C_{\rho\rho}(q,t\to \infty)}{\tilde S(q)},
 \label{6}
\end{equation}
where $f_q=0$ for all wave-vectors indicates the ergodic state which is a liquid state while $f_q>0$  indicates the non-ergodic (i.e glass) state.

Using Eq.(\ref{6}) in  Eq.(\ref{1}), the following equation for NEP can be obtained. 

\begin{equation}
f_q=\frac{M(q,t\to\infty)}{M(q,t\to\infty)+{\Omega(q)^2}},
\label{7}
\end{equation}
where $M(q,t\to\infty)$ is evaluated using Eq. (\ref{5}). $\tilde C_{\rho\rho}$ being constant for $t\to\infty$ the derivative terms  vanish and the long time
 limit of memory function is given by
 \begin{equation}
 M(q,t \to\infty)= \frac{ m\beta}{8\pi^2q^3n}\int_0^{\infty} dk k\int_{|q-k|}^{q+k} d\kappa \kappa V^2_{q,k,\kappa}f_{\kappa}f_k\tilde S(\kappa)\tilde S(k).
 \label{9}
 \end{equation}
 Eqs. (\ref{7}) and (\ref{9}) should be solved in self-consistent manner to calculate the nonergodic parameter. This parameter determines the critical 
 conditions for the liquid-glass transition at different  degrees of quantum fluctuation. The NEP was analyzed in Refs. \cite{Markland2011} to predict 
 reentrant behavior of liquid-glass transition as the quantumness is varied. 
 
\section{Results}
\begin{figure}
 \begin{center}
 \begin{center}
 \includegraphics[width=11 cm,height=8 cm]{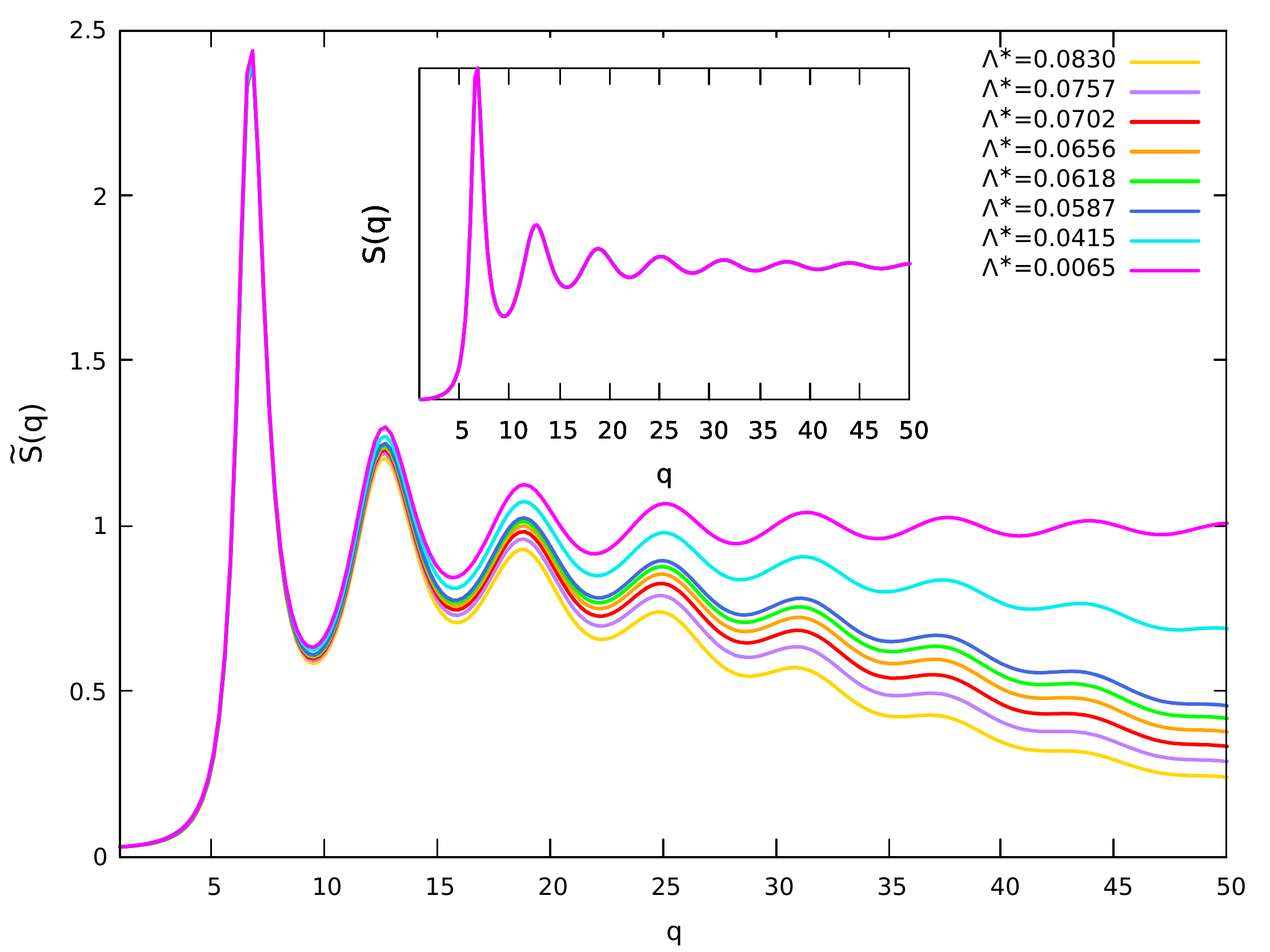}
\end{center}
 \end{center}
\caption{Kubo-transformed structure factor is shown for $\eta=0.445$ for different quantum fluctuations. With increasing $\Lambda^*$ values, $\tilde S(q)$ 
at large wave vector deviates more from unity. Inset shows the quantum structure factor for different $\Lambda^*$ values with no visible difference for 
different $\Lambda^*$. $q$ is in the units of $\sigma^{-1}$.}
\label{fig1}
\end{figure}

The self-consistent approach of QMCT requires the static structure factor and Kubo transformed structure factor as input. RISM equations with 
PY closure were used to generate the inputs for a single component hard sphere (HS) system \cite{Markland2012}. The degree of quantumness is 
measured through the dimensionless parameter $\Lambda^*$ defined above. By decreasing the mass or temperature of the system we can control 
the value of $\Lambda^*$; greater value of $\Lambda^*$ indicates higher quantum fluctuations in the system. Figure (\ref{fig1}) shows the Kubo 
transformed structure factor for different values of $\Lambda^*$ at volume fraction $\eta=0.445$ ( The classical HS system shows transition at 
critical $\eta_c=0.5089$). Inset shows the quantum structure factor at $\eta=0.445$ for different values of $\Lambda^*$. The quantum structure factor 
$S(q)$ is not affected much as $\Lambda^*$ is varied, but the Kubo-transformed structure factor $\tilde S(q)$  shows significant dependence on 
$\Lambda^*$, especially at larger $q$ values where it tends to decrease with increasing $q$, which enhances as $\Lambda^*$ is increased. 
As large value of $q$ corresponds to short length-scales, the structure factor at large $q$ gives information about self-part of the density correlation, 
which  by definition is unity for $S(q)$, as seen in the inset of the figure. Kubo-transformed structure factor corresponds to the structure factor of a 
ring-polymer consisting of infinite number of beads \cite{Chandler1981}.  Upon increasing the quantumness,  the polymer becomes more flexible 
leading to more uncertainty in the position of the quantum particle which leads to smaller values for Kubo correlation as $q$ is increased.

\begin{figure}
\begin{center}
 \includegraphics[width=10 cm,height=7 cm]{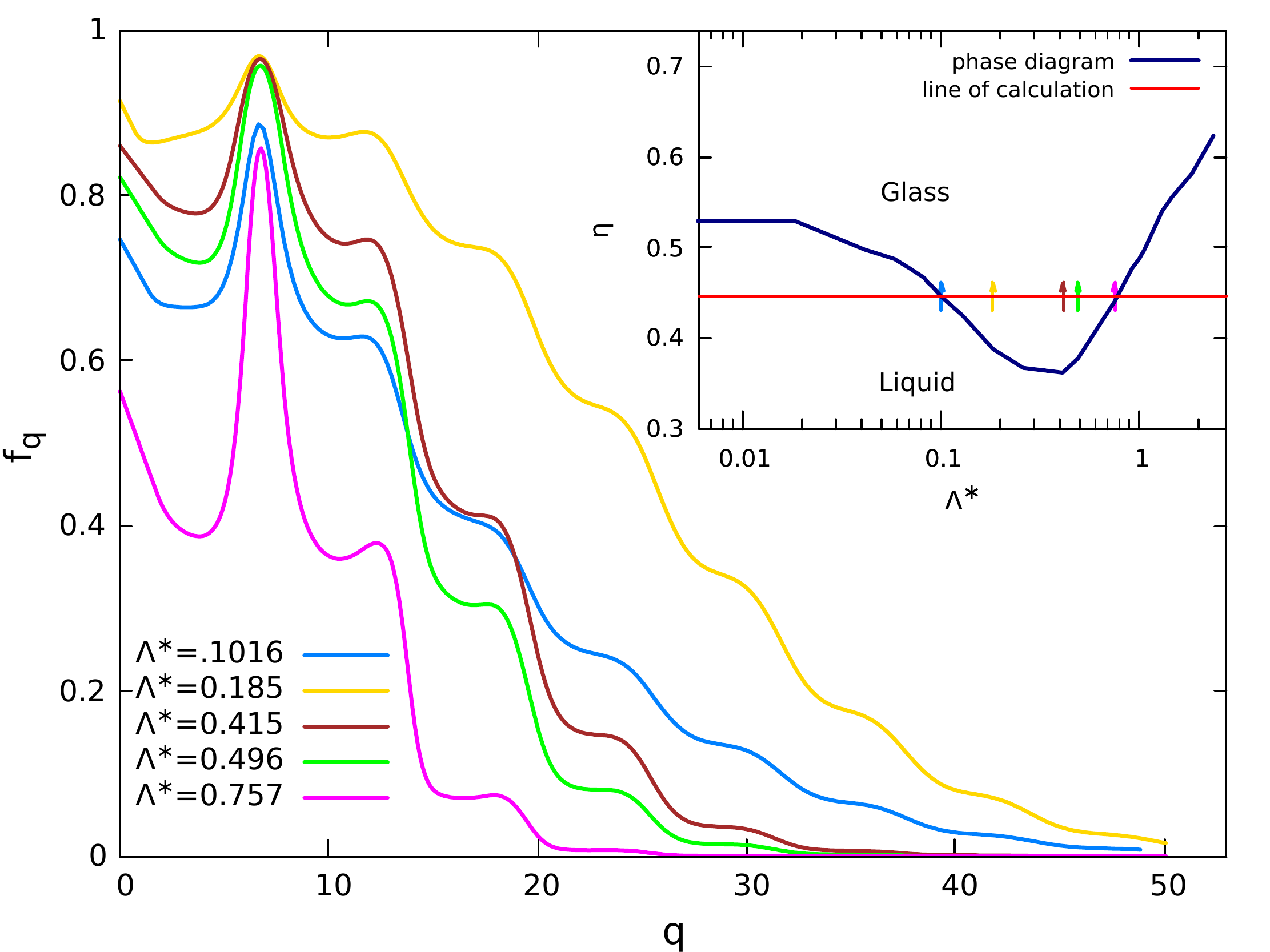}
\end{center}
\caption{Non-ergodicity parameter at $\eta=0.445$ as function of $q$ is plotted for varying degree of quantum fluctuation. Inset shows phase diagram for 
liquid-glass transition with increasing quantum fluctuation. The red line is at $\eta=0.445$ along which the dynamics and frequency-dependent specific heat 
is calculated. The positions of the NEPs in $\Lambda^*$ are shown  on the red line with  small arrows.}
\label{f2}
\end{figure}
 
 Using $S(q)$ and $\tilde S(q)$ as inputs, we solve Eqs.(\ref{7}) and (\ref{9}) self-consistently to obtain NEPs and the critical density for different $\Lambda^*$ 
 values. Some results for NEP are shown in Fig. (\ref{f2})  at $\eta=0.445$ for various values of $\Lambda^*$. In the inset we reproduce the results of 
 Ref.\cite{Markland2011} for the reentrant behavior of liquid-glass transition as $\Lambda^*$ is varied . The NEP in blue color corresponds to $\Lambda^*=0.1016$ 
 which is liquid-glass transition point for $\eta=0.445$ marked with blue arrow in the inset. If $\Lambda^*$ is increased along the line of $\eta=0.445$,  
 the first peak of $f_q$ rises up to $\Lambda^*=0.415$, on further increasing  $\Lambda^*$ the peak height of $f_q$ starts decreasing. 
 Beyond $\Lambda^*=0.757$, the $f_q$ is zero for all $q$ values and the system reenters the liquid phase. Increasing $\Lambda^*$ 
 further, the transition point shifts to lower values of density and beyond $\Lambda^* \approx 0.3$, the critical density starts to 
 move towards the classical value. We find that beyond $\Lambda^*> 1$, the 
 \begin{figure}
\begin{center}
\begin{center}
 \includegraphics[width=10 cm,height=8 cm]{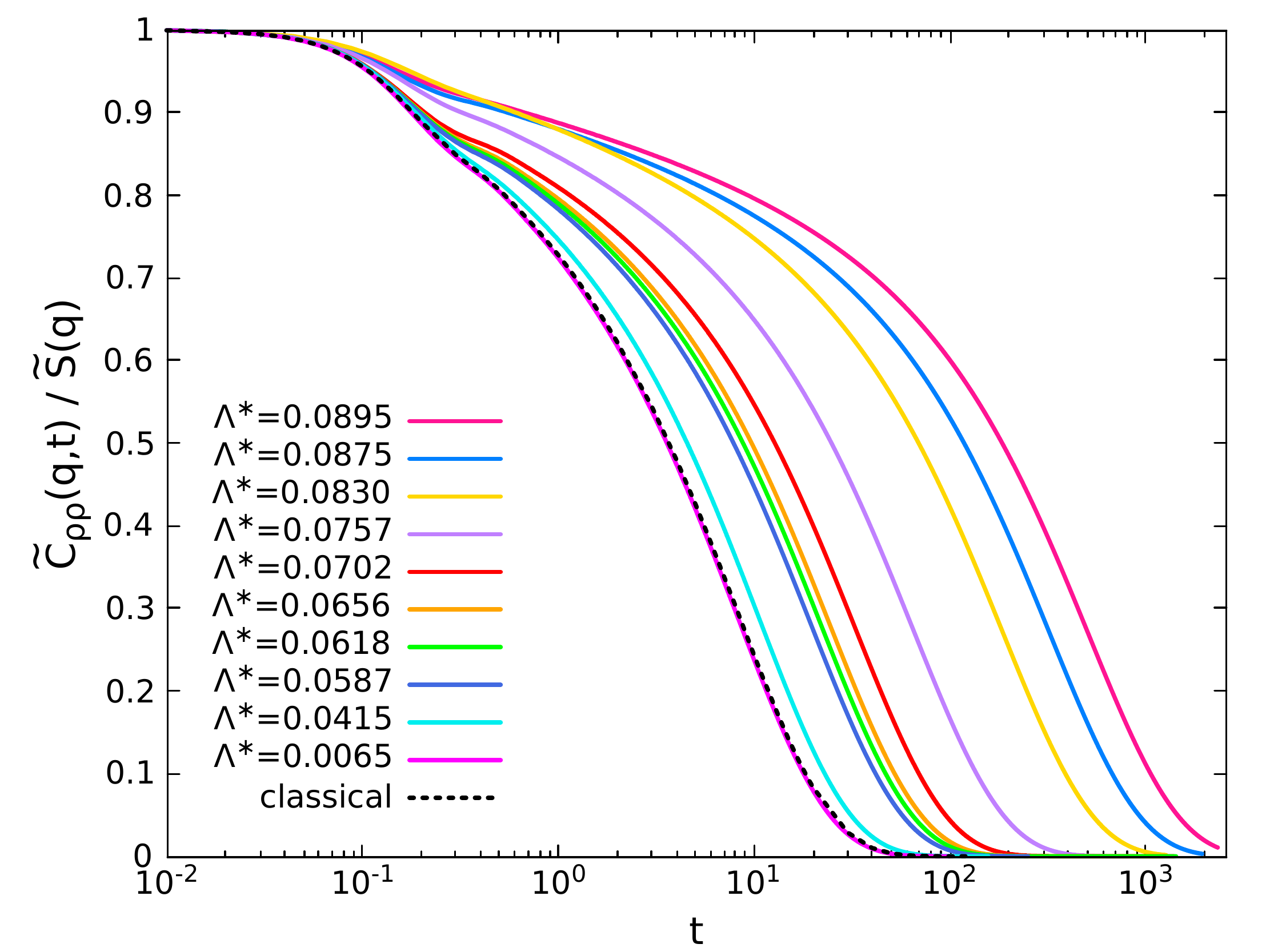}
\end{center}
\end{center}
\caption{ Dynamics of the density correlation is plotted at volume fraction $\eta=0.445$ and $q=7.11$ with different degrees of quantum fluctuation. 
The black dotted curve represents the classical dynamics. Dynamics slows down with increase in quantum fluctuation. }
\label{f3}
\end{figure}
 transition occurs at densities higher than the classical value. For $\Lambda^*=2.39$ the liquid-glass transition occurs at $\eta_c=0.623$ which is considerably 
 high compared to classical critical density and close to the value $\eta=0.74$ for closed pack system (e.g face center cubic lattice). We realize that at such high 
 densities, the current results may not be very reliable as PY-approximation used to generate the static structure may fail. However,  due to quantum effects 
 it should be possible to go to higher densities retaining the supercooled liquid phase. The phase diagram suggests the existence of two different kinds of 
 liquids, one at the region of lower $\Lambda^*$ dominated by classical interaction and the other in the region of higher $\Lambda^*$ dominated by 
 enhanced quantum tunneling effect. We are able to calculate the dynamics of the supercooled liquid of the first kind of liquid in low $\Lambda^*$ 
 region ($\Lambda^*\leq 0.092$), along the red curve at $\eta=0.445$ in the inset. This limitation is due to the use of perturbative calculation as 
 described above. It will be interesting to explore the dynamic and structural differences in the liquids in the two extreme values of quantumness. 
 This will require going beyond the perturbative approach and will be discussed elsewhere.

\begin{figure}
 \begin{center}
 \includegraphics[width=15 cm,height=10 cm]{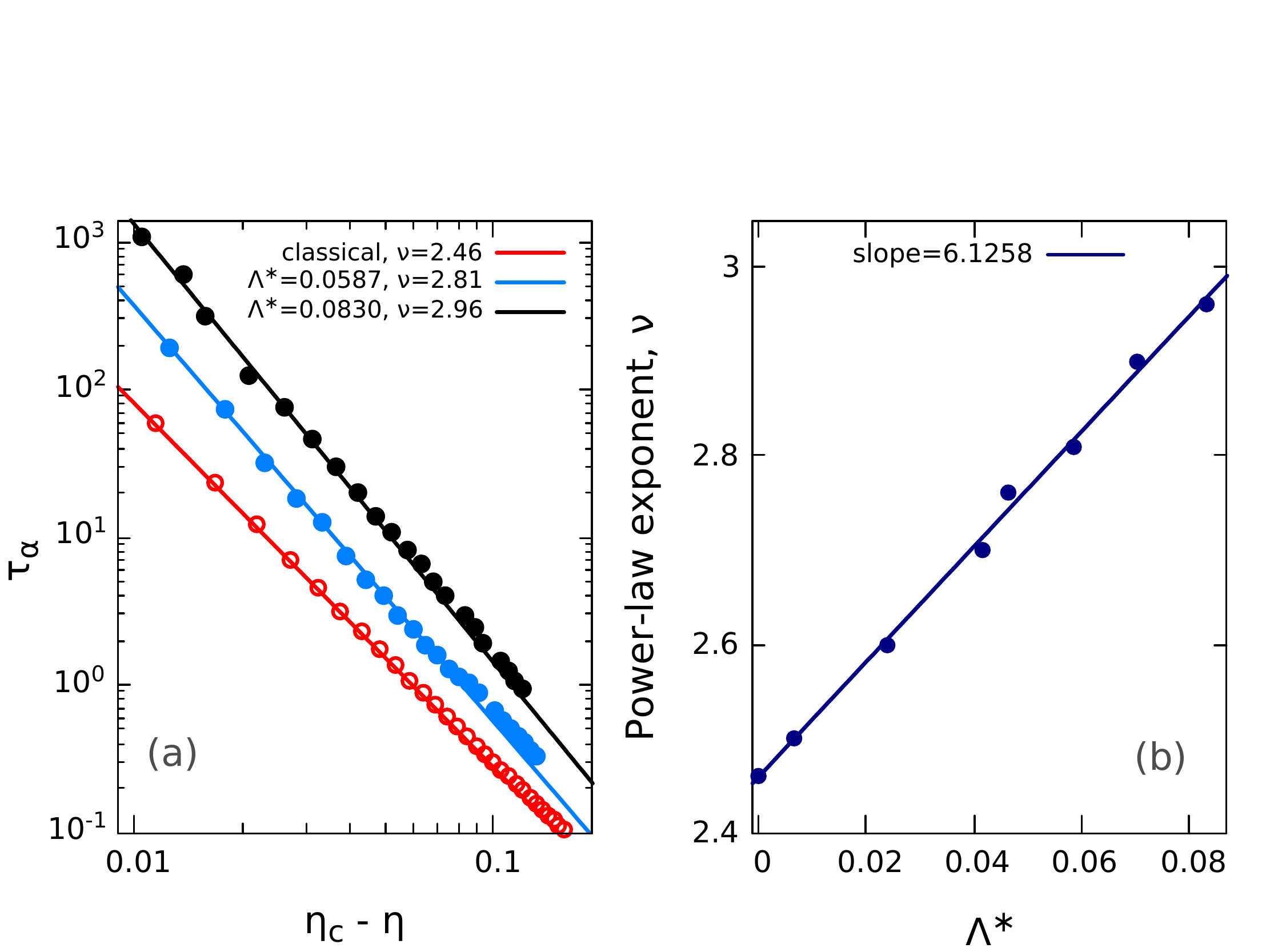}
\end{center}
\caption{ (a) Increase in the relaxation time, $t_{\alpha}$, as $\eta$ is increased at fixed $\Lambda^*$. 
$\eta_c$ is the critical value obtained from QMCT. Dots represent the QMCT results and the solid black and the
blue curves show  power-law fits, and the red curve with void circles is the classical result. 
The red and the blue data points are shifted vertically for clarity by an amount $-\mbox{ln}(7)$ and $-\mbox{ln}(2.4)$, respectively.
(b) Variation in the power-law exponent $\nu$  as the quantumness is increased.}
\label{f4}
\end{figure}

In Fig. (\ref{f3}) the normalized Kubo density correlation is plotted against time, $t$ (in units of $\tau$), for $q=7.11$ 
(wave vector around the first peak of structure factor) and $\eta=0.445$. The black dotted curve represents the dynamics of classical hard-sphere 
system and the solid curves represent the dynamics of quantum supercooled liquid. Results for $\Lambda^*=0.0065$ matches with the classical 
dynamics, and represents classical limit of the quantum system.  Like classical density correlation the Kubo correlation also shows two-step 
relaxation, $\beta-$relaxation (the plateau region) governed by power-law decay, followed by $\alpha-$relaxation (tail region)  dictated by 
stretched exponential decay. As the quantum fluctuation increases, time taken for the correlation to decay gradually increases. 
Increase in $\Lambda^*$ in the lower region of quantumness indicates increased dimensions 
of the ring-polymer thereby the particles (each ring-polymer) experience more cage effect (jammed state) by their neighboring particles resulting 
in slower relaxation.

\begin{figure}
\begin{center}
 \includegraphics[width=10 cm,height=8 cm]{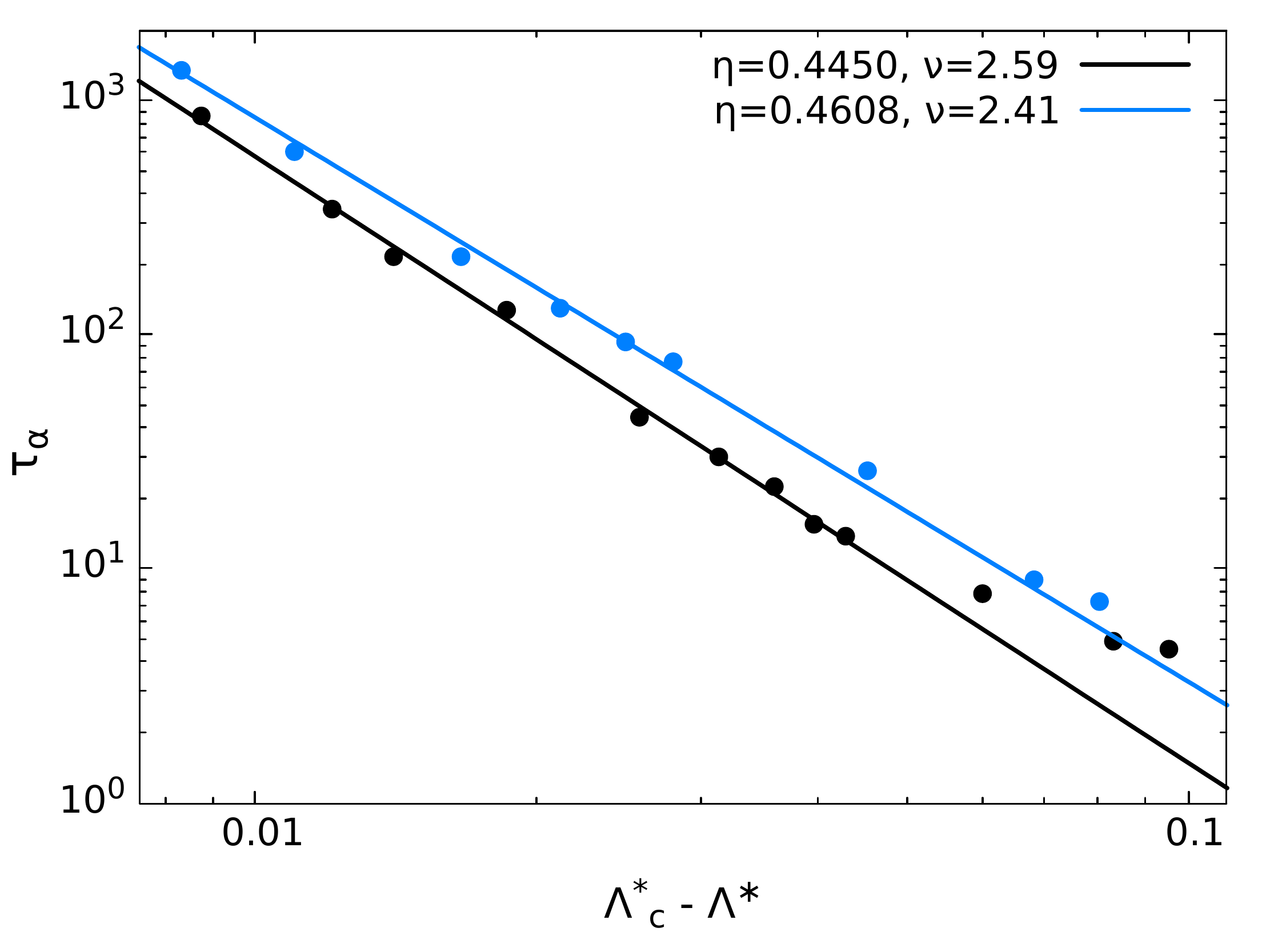}
\end{center}
\caption{Increase in the relaxation time as $\Lambda^*$  is increased at fixed densities. $\Lambda^*_c$ is 
the critical value obtained from QMCT. Dots represent the QMCT results and the solid black and the
blue curves show  power-law fit}
\label{f5}
\end{figure}

In Fig. (\ref{f4}) we show divergence in the relaxation time, $t_{\alpha}$, as critical value
of $\eta$ is approached. The relaxation time is approximated with the time where the correlation 
function decays to $1/e \approx 0.367879$ of its initial value.
The QMCT data for the relaxation time is fitted with power-law $t_{\alpha}=t_0\left(\frac{1}{\eta_c-\eta}\right)^{\nu}$  with $t_0$ and $\nu$ being the fit parameters. 
The critical values of $\eta_c$ are obtained from NEP calculation from QMCT. In Fig. (\ref{f4}a), we present results for 
the increase in the relaxation time as the density is increased, keeping $\Lambda^*$ fixed. The blue and the black 
dots correspond to $\Lambda^*=0.0587$ and $0.0830$, respectively, for which the critical densities 
are $\eta_c=0.483$ and $0.466$, respectively. The two power-law fits give different exponents. For 
$\Lambda^*=0.0587$, $\nu=2.81$ while for $\Lambda^*=0.0830$, $\nu=2.96$. 
It is known that for the classical HS \cite{PhysRevE.55.7153} and Lennard-Jones liquids\cite{PhysRevLett.73.1376, PhysRevE.52.4134}, 
 the relaxation time increases near the transition point following power-law behavior with $\nu= 2.46$  and $\nu \sim 2.5$, respectively, as 
 the (HS) density is increased or the (LJ) temperature is decreased.
 For comparison, we also show results for the classical ($\Lambda^*=0$) case (shown with open circles in Fig. (\ref{f4}a)) for which $\nu=2.46$.
The above results  indicate that, similar to the classical liquid, the relaxation time in quantum liquid also follows power-law divergence quite well 
as the density is increased,  although the power-law exponent, $\nu$, varies with the quantumness. 
To explore it further, we compute the power-law exponent for different values of $\Lambda^*$. 
The exponent increases almost linearly as $\Lambda^*$ increases from zero. This is depicted in Fig. (\ref{f4}b).

We next discuss divergence in the relaxation time, $t_{\alpha}$, as critical value
of $\Lambda^*$ is approached at fixed density. The variation in the relaxation time with $\Lambda^*$ is shown in Fig. (\ref{f5}).
The data is fitted with power-law $t_{\alpha}=t_0\left(\frac{1}{\Lambda^*_c-\Lambda^*}\right)^{\nu}$ . 
The critical values of $\Lambda^*_c$ are obtained from NEP calculation from QMCT. For $\eta=0.445$ and $0.4608$, 
 $\Lambda^*_c= 0.1016$ and $0.08675$ , respectively. 
The black  and the blue circles show QMCT data and the lines show power-law fits. We find that for 
$\eta=0.445$, the power-law exponent $\nu=2.59$ whereas for higher density, $\eta=0.4608$, the power-law 
exponent decreases to $2.41$.

Note that in the moderate quantum regime,  the critical value of $\Lambda^*$ is higher for lower densities (see inset in Fig. (\ref{f2})). 
Thus in Fig. (\ref{f5}), data for the smaller density corresponds to the higher quantumness compared to the data for the higher density. 
Since the power-law exponent is greater for larger $\Lambda^*$, a faster divergence (larger power-law exponent) is seen for the smaller density.
\begin{figure}[h]
\begin{center}
 \includegraphics[width=10 cm ,height=8 cm]{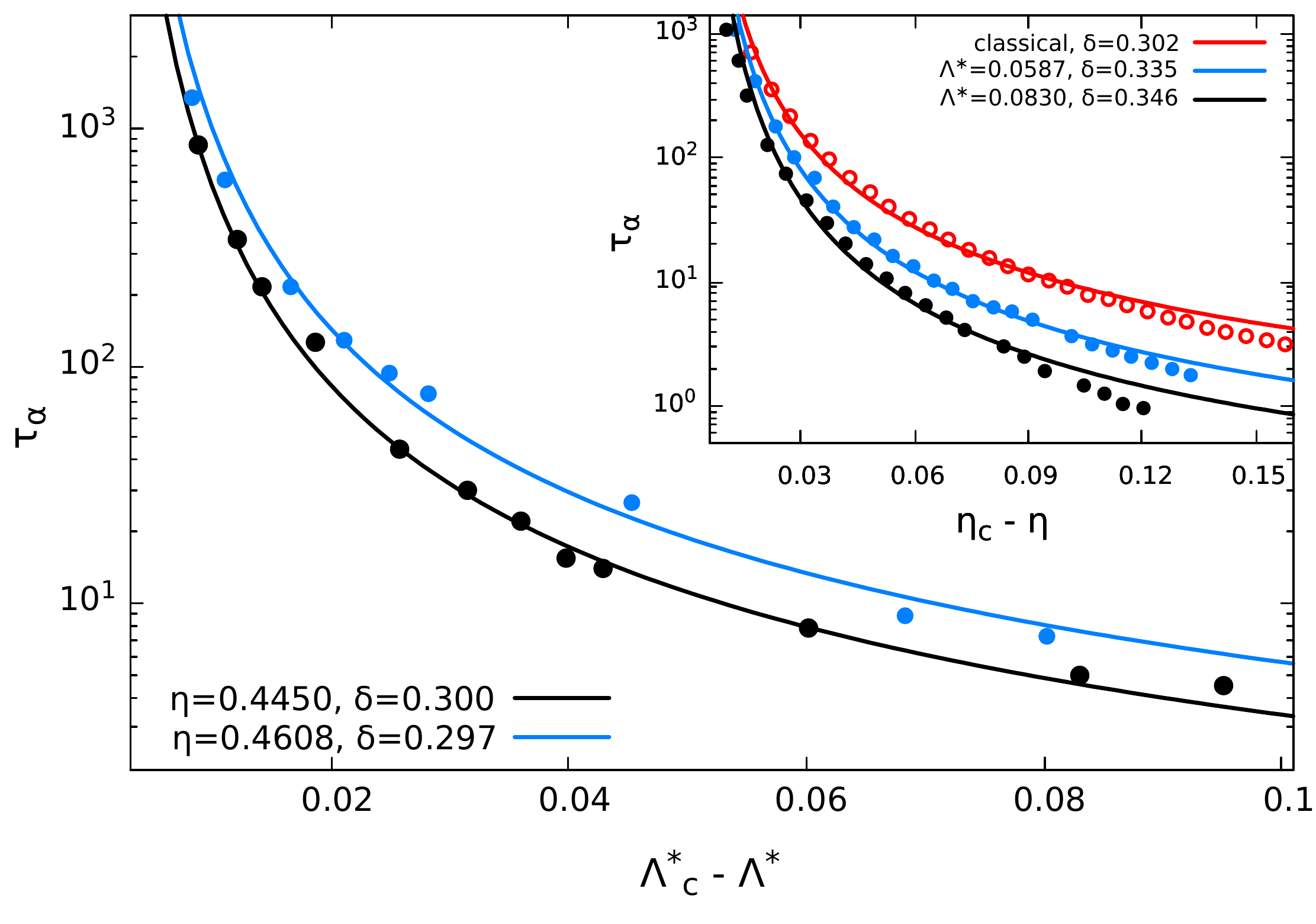}
\end{center}
\caption{ Vogel–Fulcher–Tammann (VFT)-like fit  
$t_{\alpha}=t_0\mbox{exp}\left[\frac{A}{(\Lambda_c^*-\Lambda^*)^{\delta}}\right]$  to the relaxation time, $t_{\alpha}$
as $\Lambda^*$ is increased at fixed $\eta$. Inset shows VFT fit to the relaxation time as $\eta$ is varied for fixed $\Lambda^*$.
The red and the blue data points have been shifted vertically for clarity by $\mbox{ln}(4.3)$ and $\mbox{ln}(2.3)$, respectively.}
\label{f6}
\end{figure}

For the quantum case ($\Lambda^*> 0$), the power law fits well close to the transition in Fig. (\ref{f5}), but deviates over smaller values of 
the quantumness.   We find that a Vogel–Fulcher–Tammann (VFT)-like form  
$t_{\alpha}=t_0\mbox{exp}\left[\frac{A}{(\Lambda_c^*-\Lambda^*)^{\delta}}\right]$, 
can be used to describe increase in the relaxation time quite well over the entire $\Lambda^*$ range considered. 
This is shown in Fig. (\ref{f6}) for the same data points as reported in Fig. (\ref{f5}). 
For $\eta=0.445$ and $0.4608$, the VFT exponents are $\delta=0.300$ and $0.297$, respectively.  
The VFT, however, does not describe the increase in relaxation time with respect to the density $x=\eta_c-\eta$. 
For comparison, we fit VFT to the data of Fig. (\ref{f4}a). The fit is shown in the inset of Fig. (\ref{f6}). 
It is clear that the VFT fit is not as good as the power-law fit shown in Fig. (\ref{f4}a).

\section{Conclusion}
We have used quantum mode-coupling theory to study dynamics in supercooled hard sphere quantum liquids. The quantumness of the system can 
be tuned to manipulate glass-transition point in the system. For relatively high quantumness ($\Lambda^*>1$), the critical density can surpass that 
of the critical density for classical liquid and may reach close to the closed-pack density. This is because the enhanced quantum effects lead to 
pronounced tunneling to overcome classical caging effect and allow the system to remain in ergodic phase up to higher densities.  
In the moderate quantum regime ($\Lambda^*\leq 0.1$), the relaxation time, for fixed $\Lambda^*$, increases with  
density and shows power-law divergence near the critical point. This divergence becomes stronger upon increasing the quantumness ($\Lambda^*$).

The dynamical analysis presented in this work is based on a perturbative treatment of the quantum vertex function which is valid in the moderate quantum regime. 
For the long time dynamics, the derivative terms in Eq. (\ref{5}) become less important. At $t\to \infty$, all derivative terms are strictly zero and the vertex function 
simplifies and takes the form of vertex function obtained in the classical MCT. Therefore, although the absolute value of density correlation function is sensitive to 
the derivative terms, the power-law, which is expected to emerge close to the transition point and is a characteristic of long-time relaxation, is not affected 
significantly by the derivative terms.

\section*{Acknowledgements}
AD acknowledges support from University Grants Commission (UGC), India. UH acknowledges IISc support under TATA Trust Travel Fund. 
KM acknowledges support by Japan Society for the Promotion of Science (JSPS) KAKENHI
(No.~16H04034   and  20H00128).

\section*{Data Availability }
The data that support the findings of this study are available from the corresponding author upon reasonable request.

\bibliography{ref1}

\end{document}